\documentclass[%
 reprint,
 amsmath,amssymb,
 aps,
]{revtex4-1}

\usepackage{graphicx}
\usepackage{dcolumn}
\usepackage{bm}
\usepackage{amssymb}
\usepackage{amsmath,bm}
\usepackage{tabularx}
\usepackage{rotating}
\usepackage{xcolor}

\newcommand{\be}{\begin{equation}}
\newcommand{\ee}{\end{equation}}
\newcommand{\bee}{\begin{eqnarray}}
\newcommand{\eee}{\end{eqnarray}}

\newcommand{\V }[1]{\overrightarrow{#1}}

\begin{document}


\title{Inertial drag in granular media}

\author{Shivakumar Athani}
\email{shivakumar.athani@sydney.edu.au}
\author{Pierre Rognon}%
 \email{pierre.rognon@sydney.edu.au}
\affiliation{Particles and Grains Laboratory, School of Civil Engineering, The University of Sydney, Sydney, NSW 2006, Australia}

\begin{abstract}
Like in liquids, objects moving in granular materials experience a drag force. We investigate here whether and how the object acceleration affect this drag force.
The study is based on simulations of a canonical drag test, which involves vertically uplifting a plate through a granular packing with a prescribed acceleration pattern. Depending on the plate size, plate depth and acceleration pattern, results evidence a rate-independent regime and an inertial regime where the object acceleration strongly enhances the drag force. We introduce an elasto-inertial drag force model that captures the measured drag forces in these two regimes. The model is based on observed physical processes including a gradual, elasto-inertial mobilisation of grains located above the plate. These results and analysis point out fundamental differences between mobility in granular materials upon steady and unsteady loadings. \\
\end{abstract}
\maketitle

\section{Introduction}

Granular materials are comprised of inertial grains interacting via elastic and dissipative contacts. These interactions control the macroscopic mechanical behaviour of granular packings, which is typically elasto-visco-plastic. Contact network enables finite elastic deformations, while contact sliding, opening and formation enable large visco-plastic deformation \cite{andreotti2013granular}. 

These elementary mechanical properties underpin the ability to move of large objects embedded into granular packings. In Newtonian fluids, mobility of large objects is simply described by drag force models such as Stokes and turbulent drags, which relate the speed of the object to the reaction force from the fluid. In contrast, in granular packings, the mobility of an object involves at least two distinct processes: initiating and sustaining the motion. Accordingly, models were successfully developed that establish the nature of two forces: the maximum drag force $F_s$ an object initially at rest experiences when pulled through the packing, and the final drag force $F_d$ it experiences afterward, while steadily moving. 

Models predicting maximum drag -also called ultimate capacity- have long been established for quasi-static loadings. They are routinely used in the design of building foundations in granular soils such as sand \cite{randolph2011offshore,das2013earth,das2015principles,yi2012numerical,zhang2018numerical}. For vertical uplift loadings, the maximum drag $F_s$ is proportional to  the vertical hydrostatic normal stress $\sigma_h$ at the object depth:

\be \label{eq:Onset}
F_s = N_{\gamma} S \sigma_h 
\ee

\noindent where $S$ $[m^2]$ is the surface area of the object projected in the vertical direction,  $\sigma_h = \gamma_g H$ $[N/m^2]$ with $H$ [m] the object depth and $ \gamma_g$ $[N/m^3]$ the unit weight of the granular packing. $N_{\gamma}$ is a dimensionless parameter with reported values ranging from $1$ to $100$. Several studies have shown how this parameter varies with the internal friction angle $\phi$ of the packing \cite{meyerhof1968ultimate,rowe_behaviour_1982,murray_uplift_1987,merifield2006ultimate,das2013earth,kumar2008vertical}, the object shape  \cite{khatri2011effect,bhattacharya2014pullout,dyson2014pull,askari2016intrusion,giampa2018effect} and the grain size \cite{sakai1998particle, athani2017grain,costantino_starting_2008}.  The experimental method used to measure this parameter consists of uplifting the object at a constant and relatively slow velocity and monitoring the drag force during the uplift. The maximum drag $F_s$ is then defined as the maximum value of the force opposing this motion. 
At low uplift velocities, the maximum drag is rate-independent \cite{metayer2011shearing}, which is consistent with an elasto-plastic behaviour of the packing. 

The few reported tests performed at higher uplift velocities, including experimental results in dense sand using pipes \cite{hsu1993rate,tagaya1988scale} and plate anchors \cite{bychkowski2016pullout}, revealed a linear increase in maximum drag with the uplift velocity.  However, the origin of this rate-effect remains poorly understood. 

Studies focusing on the final drag force $F_d$ provide a hint toward explaining such rate effects. Final drag forces are usually measured by moving an object through a granular packing at constant speed and measuring the average force $F_d$ needed to sustain this motion. With this method, experimental and numerical results evidenced two regimes. At low speed, the drag force $F_d$ is rate-independent and captured by a model similar to (\ref{eq:Onset}). Accordingly, the drag force in this regime is referred to as \textit{frictional} drag \cite{albert1999slow,albert2000jamming,albert2001granular,gravish2010force,costantino2011low,ding2011drag}. At higher velocities, the drag force $F_d$ exhibits a quadratic increase with speed that is reminiscent of a turbulent drag \cite{percier2011lift,potiguar2013lift,takehara2010high,takehara2014high}. This results from the inertial forces developing when grains are moving from the front of the object to its back. This points out that grain inertia can contribute to hindering the motion of objects, and could possibly be at the origin of the rate effects evidenced on the maximum drag $F_s$. 

The role of grain inertia on the mobility of objects in granular materials has been further evidenced under cycling loading in \cite{athani2018mobility}. This study showed that an object could sustain an external force larger than $F_s$ for a short period of time without moving. This effect was attributed to the grain inertia near the object, which hinders the object motion on short time scales. Accordingly, one could expect that an object being initially at rest and set into motion with some acceleration could possibly experience an increased maximum and final drag force, which would arise from the inertial displacement of accelerated grains in its surrounding. However, such an effect has not been evidenced to date, and there is therefore no established model to capture it.

The purpose of this paper is to measure whether and how grain inertia impedes the mobility of large objects in granular packing. In this aim, we conducted a series of elementary mobility tests using a discrete element method.  Tests involve prescribing a vertical acceleration to a plate embedded into a granular packing, and measuring the drag force opposing that motion. Our approach is comprised of two steps:  we first empirically measure the maximum drag for different acceleration patterns, grain stiffnesses, plate sizes and plate depths; we then develop  a model capturing these measurements, based on physical processes evidenced during the uplift.

The paper is organised as follows. Section \ref{sec:method} presents the simulation method and the details of the mobility tests. Sections \ref{sec:Peak force} and \ref{sec:model} present the measured drag forces and the model we introduce to capture them. 

\begin{figure}
\centering
  \includegraphics[width=0.5\textwidth]{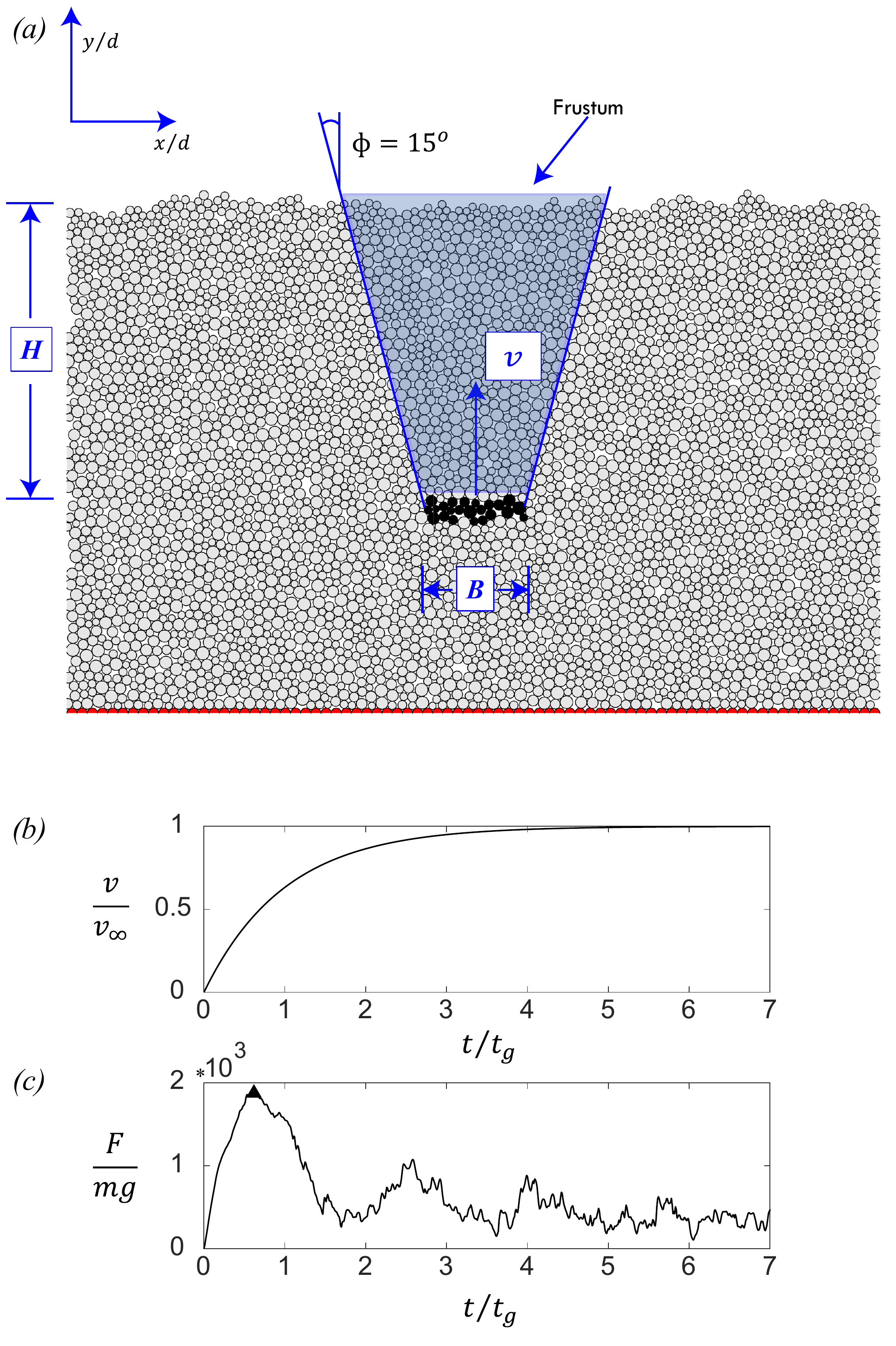}
\caption{Dynamic uplift tests. (a) Example of a system comprised of a plate (black grains) embedded in a granular material (gray grains). The domain is bidimensional and periodic in the $x$ direction. There is a layer of fixed grains (red  grains) at $y=0$.  The system width is $8B$ and the plate is placed at a distance of at least $B$ from the bottom. The blue area illustrates the frustum of grains being uplifted in quasi-static loadings. The weight of the grains in this zone corresponds to the maximum drag as per Eqs. (\ref{eq:Onset}) and (\ref{eq:Ngamma}). (b) Example of prescribed velocity of the plate, according to Eq. (\ref{eq:Vinfi}) with $\tau/t_g = 1$, showing a phase with some acceleration ($t\ll \tau$) followed by a nearly constant velocity ($t \gg \tau$). (c) Corresponding drag force $F$ during the uplift, showing a peak force $F_s$ (triangle) followed by a significant decay and some fluctuations.}
\label{fig1}     
\end{figure}

\begin{figure*}
\centering
  \includegraphics[width=\textwidth]{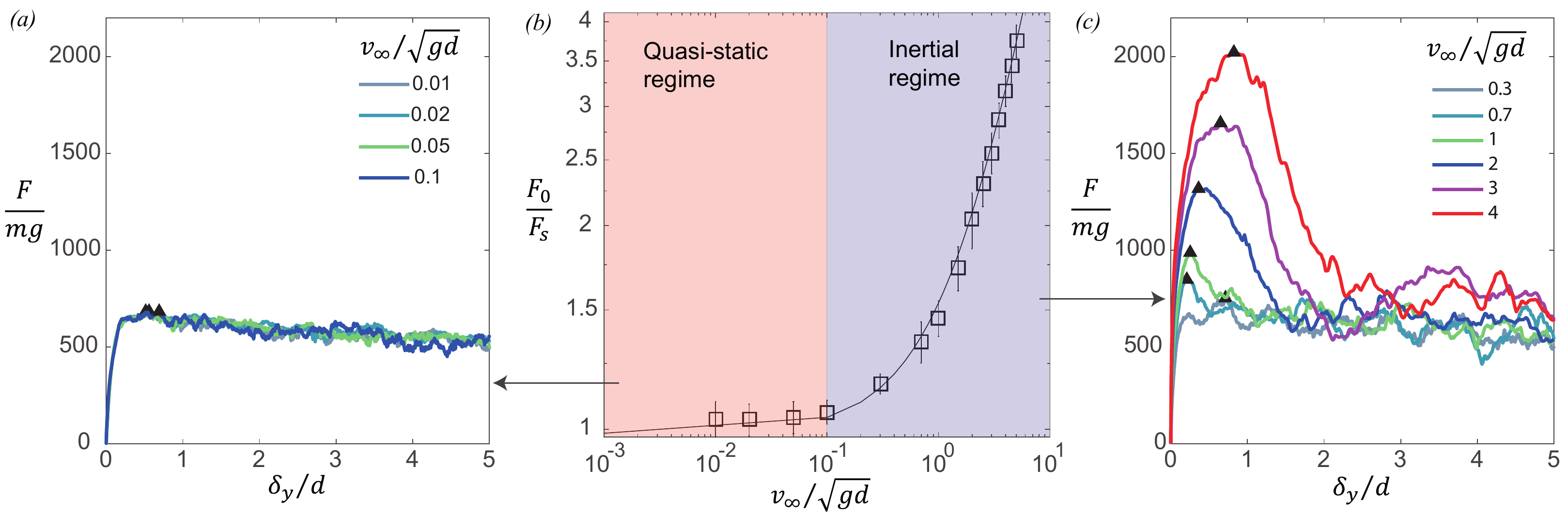}
\caption{Effect of the ultimate uplift velocity $v_{\infty}$ on the maximum drag $F_0$ for a system with $B = 10d$, $H/B = 3$ and $\tau = t_g$. Examples of drag force evolution $F$ versus plate displacement $\delta_y$ obtained (a) in the quasi-static regime ($v_{\infty} \lesssim \sqrt{gd}/10$) and (c) in the inertial regime  ($v_{\infty} \gtrsim \sqrt{gd}/10$); maximum drag forces $F_0$ are marked with a triangle. (b) Maximum drag $F_0$ measured at different velocities $v_\infty$ -  symbols and error bars show the average and standard deviation of the maximum drag obtained by repeating five similar tests with different realisations of initial packing; the black line shows the best fit of the empirical model (\ref{eq:F_v_emp}), using  $\alpha = 390 m\sqrt{g/d}$ as a fitting parameter and Eq. (\ref{eq:Ngamma}) for the quasi-static maximum drag $F_s$.}
\label{fig2}      
\end{figure*}

\section{Simulated system}\label{sec:method}

We consider a bi-dimensional system comprised of a horizontal plate embedded in a packing of cohesionless and frictional grains. Figure \ref{fig1}a illustrates this configuration. The dynamics of the system is simulated using a discrete element method. In \cite{athani2017grain}, we used a similar method and system to investigate the quasi-static uplift capacity of plate anchors; this study showed that such 2d numerical tests qualitatively match established experimental measurements of uplift capacity in dense sand, captured by Eq. (\ref{eq:Onset}).

This section briefly presents the physical parameters of the grains, the protocol of the dynamic uplift tests, and the dimensional analysis of these tests.

\subsection{Granular material}
 
Grains are disks of mean diameter $d$ and mass $m$. A polydispersity of $d \pm 30\%$ is introduced on the grain diameter to avoid crystallisation, using a uniform distribution by number within this range. Grains are subjected to gravity.  They interact with their neighbours \textit{via} inelastic and frictional contacts characterised by a Young's modulus $ E$, a coefficient of restitution $e_r$ and coefficient of friction $\mu = 0.5$. 

There is no interstitial fluid in the pores or long range interaction. Grain translation and rotation are simulated over time using a discrete element method similar to that introduced in \cite{athani2017grain,athani2018mobility,rognon2015long}.

The plate moving through the packing is made of grains that are similar to the free grains described above. However, all plate grains move vertically at the same prescribed velocity. They do not translate horizontally or rotate. The drag force on the plate is monitored at anytime by summing up all contact forces between free grains and plate grains. This drag force therefore corresponds to the net reaction force of the granular packing, and excludes the weight of the plate. In the following, $F(t)$ indicates the vertical component of this reaction force using the following convention: positive values correspond to a reaction force oriented downward.  We checked that the horizontal component of the reaction force is always close to zero.

\subsection{Dynamic uplift tests}

Dynamic uplift tests involve preparing a dense and static packing of grains, placing a plate into it, and pulling the plate at a controlled velocity that can vary in time. Simulations are conducted in a domain that is periodic in the x-direction.  Therefore, the simulated system is an array of plates rather than a single plate. We used a domain size -corresponding to the horizontal spacing between plates- of $L = 8B$ for the tests presented in the following. We consistently observed that using larger system sizes did not affect the results, which indicates that the simulated system is representative of a single plate behaviour.   

Dense packings are formed by initially placing grains at random locations in a loose configuration, without contact. Grains then settle under the action of gravity $g$ into a denser configuration, with virtually no kinetic energy.  During this preparation, grains are subjected to a background drag force of the form $\V f_i^{drag} = -\xi \V v_i$, where $\V v_i$ is the velocity of grain $i$ and $\xi$ a drag coefficient. This background drag is introduced to limit the maximum free fall velocity of grains to $v_{max} = \frac{ mg}{\xi}$, which restricts the build up of kinetic energy during settling. Once there is virtually no kinetic energy left in the system, the resulting packing has a porosity of about $20\%$ and an internal friction angle of ${15}{^o}$. The background drag is then turned off for the uplift tests. 

The plate is created within the static dense packing by selecting free grains at a desired location and tagging them as plate-grains. This method avoids the creation of heterogeneities in the granular packing that would arise by either (i) placing a plate and pouring the grains or (ii) pushing a plate into a granular packing. Moreover, this method automatically produces a plate that is at mechanical equilibrium: the sum of the contact forces between free grains and plate grains balances the weight of the plate grains. The plate thus formed is not smooth: it features asperities of the order of the grain size. 

The dynamic uplift tests are conducted by controlling the upward displacement of the plate using a velocity/acceleration pattern characterised by two parameters: a final uplift velocity $v_{\infty}$ and an acceleration time $\tau$. At any time $t$, the vertical plate acceleration and velocity along the $y$ direction are defined by:

\bee \label{eq:Vinfi}
v(t) &=& v_{\infty} \left( 1-e^{ -\frac{t}{\tau} } \right);\\
a(t)&=& \frac{v_{\infty}}{\tau} e^{- \frac{t}{\tau}}.
\eee

\noindent The convention used for the velocity is that positive values correspond to upward motion. $t=0$ is the beginning of the uplift test, when the plate and granular packing are at rest. These exponential functions are chosen to smoothly transition from an accelerated motion at the beginning of the test ($t\lesssim \tau$) during which the average acceleration is $\frac{v_{\infty}}{\tau}$, to a steady motion with a constant uplift velocity $v_{\infty}$ afterward ($t\gg \tau$). 


\subsection{Dimensional analysis}\label{sec:units}

The simulated system is defined by a number of geometrical and physical parameters that form elementary time, force and length scales. 
In the following, we will express masses, lengths and forces in unit grain mass $m$,  diameter $d$ and weight $mg$, respectively. Accordingly, the unit time is $t_g = \sqrt{d/g}$. It represents to the time for a grain to free fall over a distance $d$ under the action of gravity. 

The mode of loading involves two elementary time scales: the acceleration time $\tau$, and the ultimate plate displacement time scale  $t_p = d/v_{\infty}$. 
The fact that grains are inertial and elastic leads another time scale that represents a binary collision time between two grains: 

\be 
t_c = \sqrt{\frac{m}{Ed}}
\ee

\noindent This time  can also be interpreted as the time required for elastic waves to travel through a distance $d$. In the following simulations, the elastic modulus of the grains is $E = 10^4$ $mg/d^2$ so that the collision time is always shorter than the gravity time: $t_c = \frac{t_g}{100}$.

The time step $dt$ of the simulations is defined as a fraction (1/20) of the shortest time scale of the system. We checked that shorter time steps did not affect the results. 

\begin{table}
\begin{tabular}{cccccc}
\noalign{\smallskip}
\hline
\noalign{\smallskip}
$B/d$ & $H/B$ & $e_r$& $E/(mg/d^2)$ & $v_{\infty}/\sqrt{gd}$ & $\tau/t_g$ \\
\hline\noalign{\smallskip}\noalign{\smallskip}
$10$  & $2$  & $0.3-0.7$& $10^4$ & $0.1 - 5$ & $0.01 - 4$ \\
$10$  & $3$  & $0.5$& $10^4$ & $0.1 - 5$ & $0.01 - 4$ \\
$10$  & $4$  & $0.5$& $10^4$ & $0.1 - 5$ & $0.01 - 4$ \\
$15$  & $2$  & $0.5$& $10^4$ & $0.1 - 5$ & $0.01 - 4$ \\
$20$  & $1$  & $0.3-0.7$ & $10^4$& $0.1 - 5$  & $0.01 - 4$\\
$20$  & $1$  & $0.5$ & $10^3$ & $0.1 - 5$ & $0.01 - 4$\\
$30$  & $1$  & $0.5$& $10^4$  & $0.1 - 5$ & $0.01 - 4$ \\
\hline
\end{tabular}
\caption{Explored range of parameters: plate width $B$, embedment ratio $H/B$, grain coefficient of restitution $e_r$ and Young's modulus $E$,  final uplift velocity $v_{\infty}$, acceleration time $\tau$. Parameters are expressed in the system of units defined in Section \ref{sec:units}. Unless otherwise specified, results shown in the following are obtained with $E=10^4$  and $e_r=0.5$.} 
\label{tab:Parameters}    
\end{table}

\begin{figure}
\centering
  \includegraphics[width=0.38\textwidth]{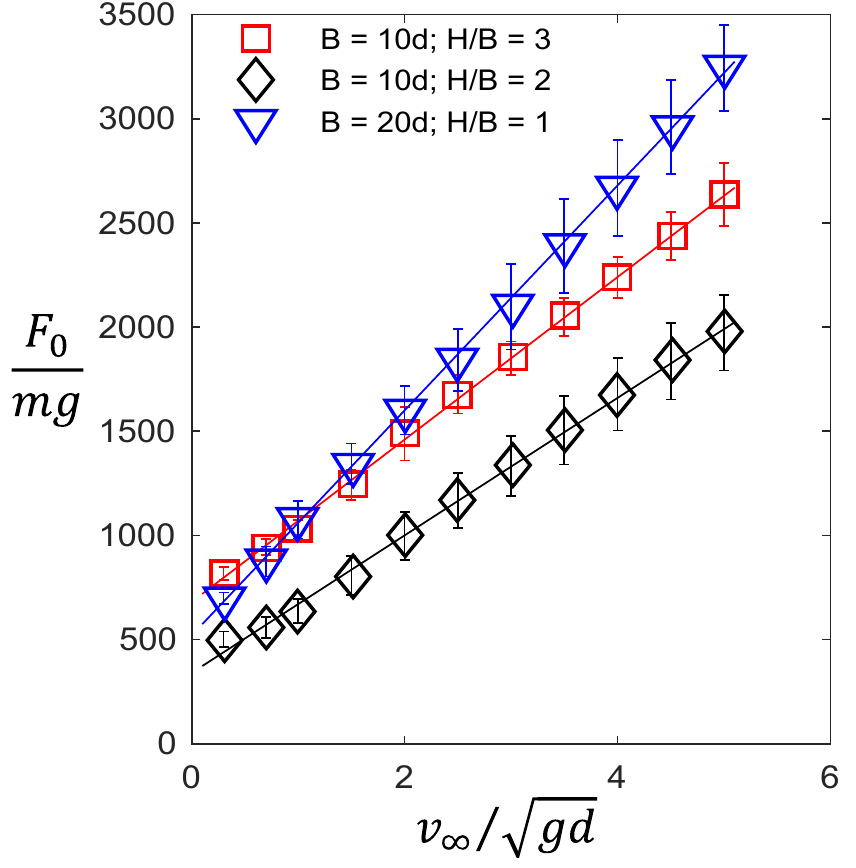}
\caption{Effect of the ultimate uplift velocity $v_\infty$ on the maximum drag $F_0$ for different plate size $B$ and plate depth $H$ ($\tau = t_g$ in all tests). Symbols and error bars show the average and standard deviation of the maximum drag obtained by repeating five similar tests with different realisations of initial packing. Lines represent the best fits of the model (\ref{eq:F_v_emp}) using $\alpha$ as a fitting parameter (best fits are obtained for $\alpha = 390, 330, 540$, respectively), while $F_s$ is given by (\ref{eq:Ngamma}).}
\label{Fig_3}      
\end{figure}

\begin{figure*}
\centering
  \includegraphics[width=0.75\textwidth]{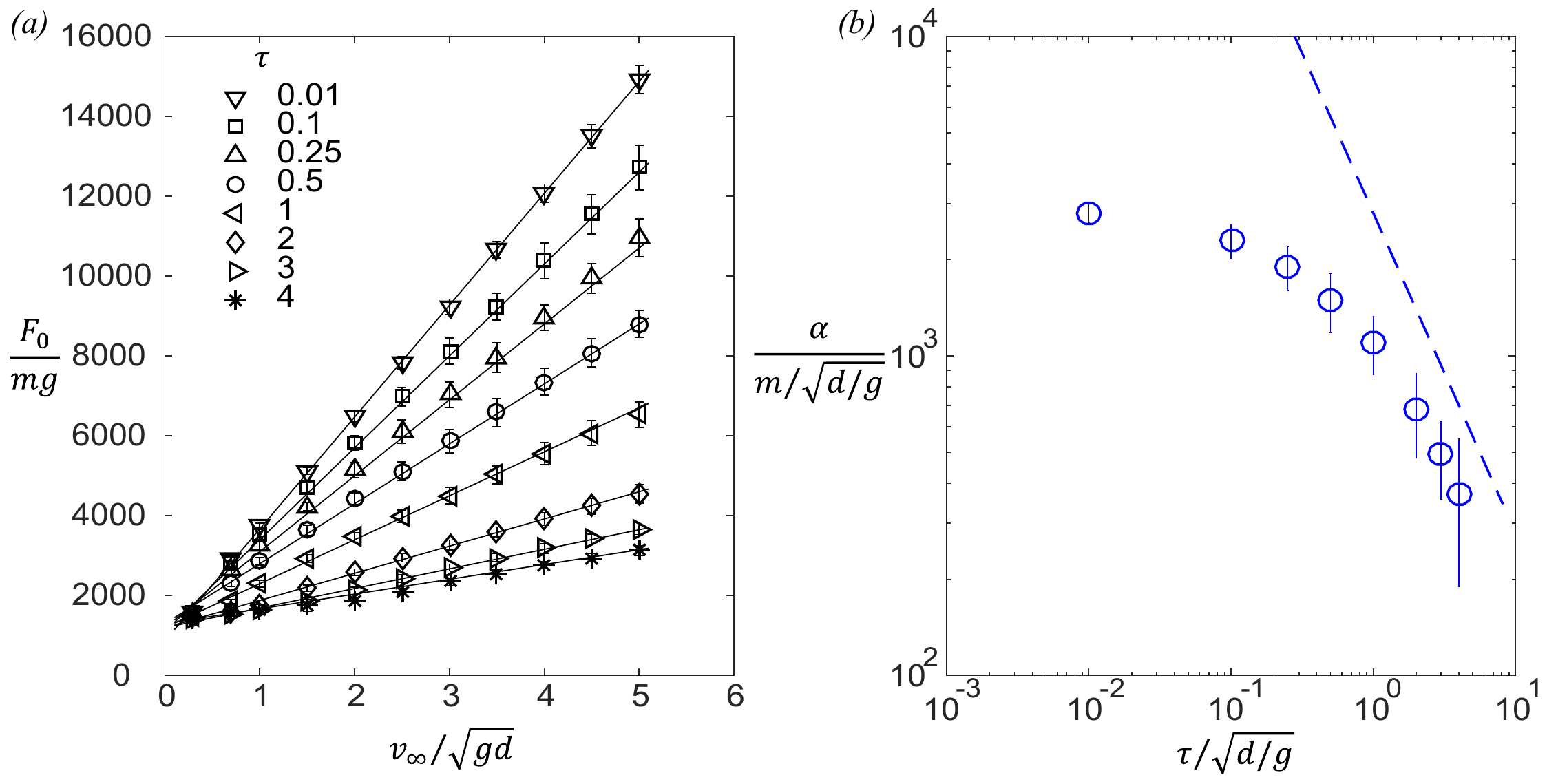}
\caption{Effect of the acceleration time $\tau$ on the uplift capacity $F_0$ for a system with $B/d = 30$, $H/B = 1$. (a) Maximum drag force $F_0$ for different ultimate uplift velocities $v_{\infty}$ and different acceleration times $\tau$. Symbols an error bars show the average and standard deviation of $F_0$ obtained on a series of five tests with different realisations of the initial packing. Lines represent the best fit of Eq. (\ref{eq:F_v_emp}) using $\alpha$ as a fitting parameter and fixing $F_s$ as per Eq. (\ref{eq:Ngamma}). (b) Values of $\alpha$ obtained with this fitting procedure. The dashed line represents a power law with an exponent $-1$ for visual reference.
}
\label{Fig_4}      
\end{figure*}

\begin{figure}
\centering
  \includegraphics[width=0.43\textwidth]{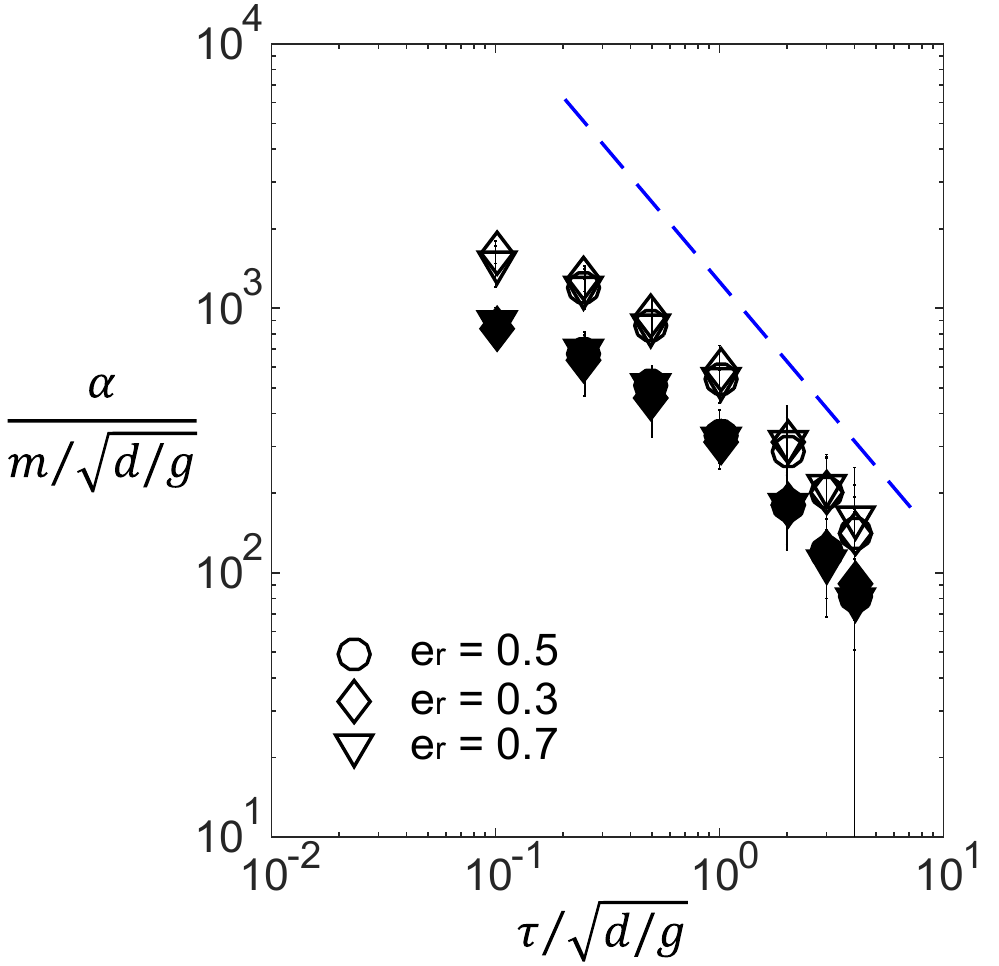}
\caption{Effect of the inter granular coefficient of restitution $e_r$ on the maximum drag $F_0$. Slope $\alpha$ measured by fitting numerical results of $F_0(v_\infty)$ by Eq. (\ref{eq:F_v_emp}), following the procedure introduced on figure \ref{Fig_4}. Open and filled symbols correspond to systems with $B/d = 20$ \& $H/B = 1$ and with $B/d = 10$ \& $H/B = 2$, respectively. The dashed line represents a power law with an exponent $-1$ for visual reference.
}
\label{Fig_5}      
\end{figure}

\begin{figure*}
\centering
  \includegraphics[width=\textwidth]{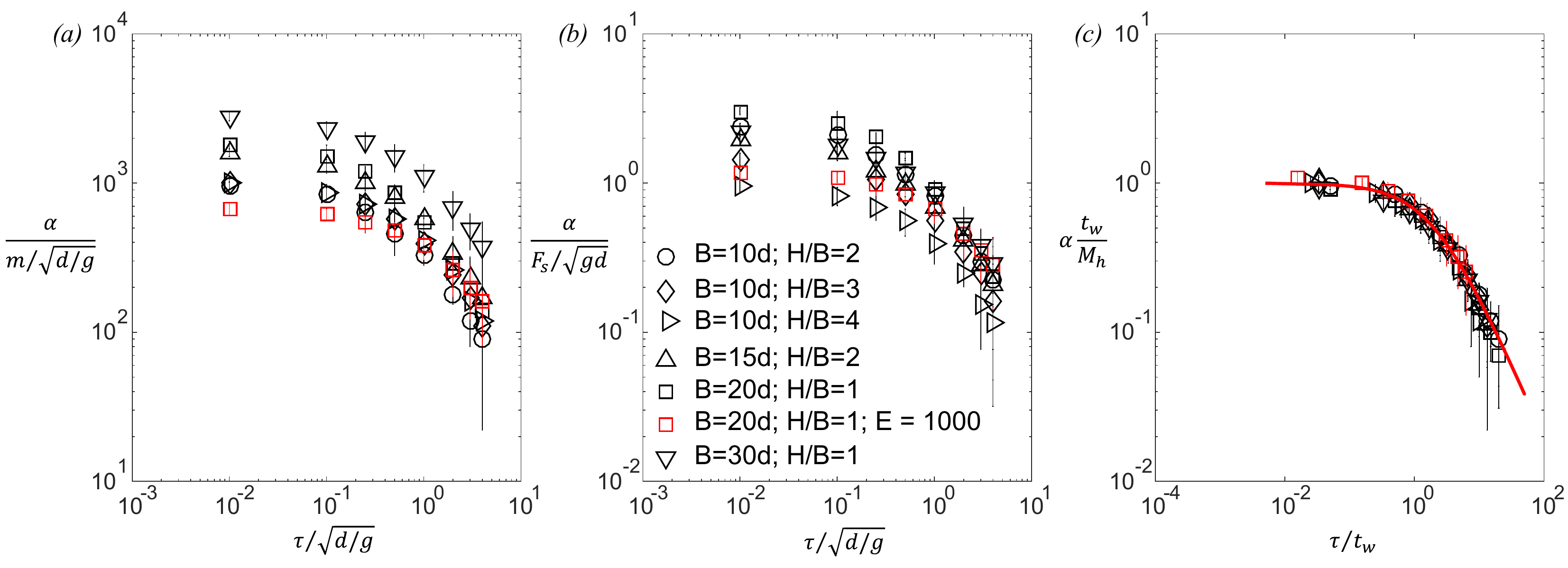}
\caption{Scaling of the coefficient $\alpha$ for different combinations of plate width $B$, plate depth $H$, acceleration times $\tau$ and grain Young's modulus (see legend in (b)). Values of $\alpha$ are obtained by fitting the measured values of $F_0(v_\infty)$ by Eq. (\ref{eq:F_v_emp}).  Black/red symbols correspond to grain Young's modulus $E = 10^4$ and $10^3$ $mg/d^2$, respectively. (a) Non-normalised values of $\alpha(\tau)$. (b) Normalisation attempt using the quasi-static maximum drag force $F_s = F_0(v_\infty=0)$ and the free fall velocity $\sqrt{gd}$. (c) Successful normalisation using the elastic wave propagation time $t_w$ and the hydrostatic mass $M_h$ defined in (\ref{eq:twave}) and (\ref{eq:Mh}). The red line represents the best fit of the elasto-inertial drag model  defined in (\ref{eq:alpha_predict}), which is obtained for $\beta=2$.}
\label{Fig_6}      
\end{figure*}

\section{Measured maximum drag force $F_s$}\label{sec:Peak force}

Figures \ref{fig1}b,c show the results of a dynamic uplift test performed with $B/d= 10$, $H/B=3$, $\tau/t_g = 1 $ and $v_\infty = 3 \sqrt{dg}$). The drag force $F$ first increases to a maximum and then sharply decreases. Similar tests have been conducted with plates of different width $B$, embedded at different depth $H$ and with different acceleration parameters $\tau$ and $v_{\infty}$. Table \ref{tab:Parameters} summarises the explored range of parameters.  All  tests produced drag force evolutions qualitatively similar to that presented on figure \ref{fig1}c, albeit with different values of the maximum drag force. In the following, we refer to the maximum drag force as $F_0$, to distinguish it from the quasi-static maximum drag force $F_s$ defined in (\ref{eq:Onset}). In this section, we seek to empirically establish how $F_0$ depends on the plate size, plate embedment, and on the acceleration parameters $\tau$ and $v_{\infty}$. The physical origin of these dependencies will be discussed in the next section. 

\subsection{Quasi-static \& inertial regimes}\label{sec:Rate Indi}

Figure \ref{fig2} shows the effect of varying the ultimate velocity $v_\infty$ on the drag force. All these tests are performed with a plate size $B=10d$, a plate depth $H=3B$ and an acceleration time $\tau = t_g$. The only parameter varying from test to test is the ultimate uplift velocity $v_\infty$. Results evidence a rate-independent regime at low velocities ($v_{\infty} \lesssim \sqrt{gd}/10$), where the maximum drag $F_0$ does not significantly depend on the rate of pull.  At larger velocities, results indicate a rate-dependent regime where the maximum drag increases approximately linearly with the ultimate velocity.  We refer to these regimes as quasi-static regime and inertial regime, respectively. As in \cite{athani2017grain}, we observed that the quasi-static maximum drag is given by Eq. (\ref{eq:Onset}) with:

\be \label{eq:Ngamma}
N_{\gamma} \approx 1+\frac{H}{B} \tan(\phi)
\ee

\noindent where $\phi\approx 15^\circ$ is the internal friction coefficient of the packing. This corresponds to the failure mode illustrated on figure \ref{fig1}a, whereby a frustum of grains is being uplifted by the plate. The quasi-static limit of the maximum drag corresponds to the weight of this frustum of grains. 

We propose to capture the maximum drag force in both the quasi-static and inertial regimes by the following linear function:

\be \label{eq:F_v_emp}
F_0 \approx F_s + \alpha v_\infty
\ee

\noindent where $\alpha$ is a coefficient with a dimension force per unit velocity, which does not dependent on $v_\infty$. Figure \ref{fig2} shows how this function fits the measured maximum drag $F_0$ in both the quasi-static and inertial regimes using $\alpha$ as a fitting parameter and fixing $F_s$ as per (\ref{eq:Ngamma}). Figure \ref{Fig_3} further indicates that this linear model captures the maximum drag forces measured with different embedment and plate size, and indicates that the coefficient $\alpha$ depends on these parameters.  

This observed linear increase of the maximum drag with the uplift velocity is consistent with previous experimental observations in dense sand using pipes \cite{hsu1993rate,tagaya1988scale} and plate anchors \cite{bychkowski2016pullout}.

\subsection{Effect of the acceleration time $\tau$} 

Figure \ref{Fig_4} shows the effect of the acceleration time $\tau$ on the maximum drag for a plate of size $B=30d$ and embedment $H=B$. The linear increase (\ref{eq:F_v_emp}) is recovered for all acceleration times. 
However, the value of the acceleration time $\tau$ strongly influences the parameter $\alpha$. For large values of $\tau$ ($\tau \gtrsim 0.1 t_g$), $\alpha$ appears to be inversely proportional to the acceleration time: 

\be 
\alpha \propto \tau^{-1} \text{ for } \tau \gtrsim 0.1 t_g
\ee 

\noindent In contrast, the parameter $\alpha$ seemingly reaches a maximum and plateaus for small acceleration times  ($\tau \lesssim 0.1 t_g$).

Figure \ref{Fig_5} shows that this dependency is also observed for systems with different values of inter granular coefficient of restitution $e_r$, plate width $B$ and embedment $H$. The coefficient $\alpha$ is not affected by the value of the inter-granular coefficient of restitution, which controls normal energy dissipation at the contact level. 

Lastly, figure \ref{Fig_6}a shows the effect of grain stiffness on the coefficient $\alpha$, by comparing systems with different values of Young's modulus $E$.  All systems lead to a qualitatively similar function $\alpha(\tau)$, including a plateau at low values of $\tau$ and an inverse power law at large values of $\tau$. However, parameters including the plate depth $H$, the plate width $B$ and the grain Young's modulus appear to quantitatively affect the value of the plateau and the value of the power law pre-factor.

\section{Dynamic drag model}\label{sec:model}

This section seeks to establish the physical origins of the maximum drag force $F_0$ as a way to explain its dependencies with the acceleration parameters. As a starting point, we detail the established process underpinning the quasi-static maximum drag $F_s$. We then introduce an elasto-inertial drag model to account for the influence of the acceleration parameters $v_\infty$ and $\tau$.

\subsection{Quasi-static uplift capacity}

The maximum drag force experienced by a plate being moved infinitely slowly corresponds to the weight of the grains it lifts up. Uplifted grains are not strictly limited to the column above the plate. They include grains enclosed in a frustum as illustrated on figure \ref{fig1}a, which geometry depends on the internal friction angle of the packing. The corresponding mass $M_s$ is (in 2d, considering a unit depth $d$ in the third dimension):

\bee 
M_s &=& M_h N_\gamma,\label{eq:Ms}\\
M_h&=& \rho BHd . \label{eq:Mh}
\eee

\noindent $N_\gamma$, given by Eq. (\ref{eq:Ngamma}), accounts for the shape of the frustum. In our system ($1 \leqslant H/B \leqslant 3$, $\theta \approx 15^\circ$), values of $N_\gamma$ range from $1.3$ to $2.1$.
$M_h$ is the mass of the grains located above the plate, and $\rho$ is the density of the packing. This process explains the quasi-static uplift force $F_s = M_s g$, which is rate-independent. While it does not account for the observed rate effects, it does point out that moving the plate requires moving some inertial grain in the packing, and therefore involves some inertia.

\begin{figure}
\centering
  \includegraphics[width=0.5\textwidth]{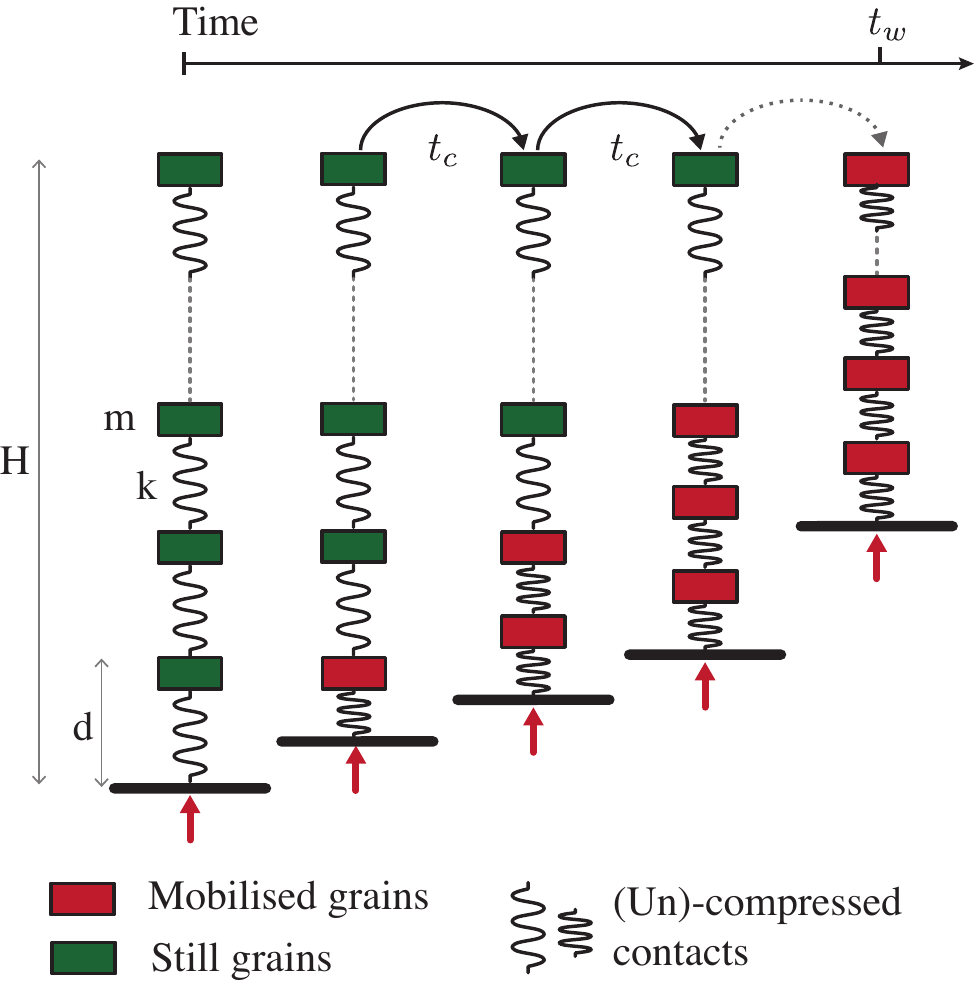}
\caption{Illustration of the elasto-inertial process controlling the inertial drag force. Rectangles represents layers of grains situated above the plate (black line), that are gradually mobilised as the elasto-inertial stress wave propagates toward the surface.
}
\label{fig7}      
\end{figure}

\subsection{Elasto-inertial drag}

We infer that an accelerating plate would be resisted by two forces: the quasi-static drag mentioned above and an inertial drag $F_i$ resulting from the grains acceleration in the packing. Accordingly, we express the maximum drag as:

\bee \label{eq:model}
 F_0  &=& F_s+ F_i \label{eq:F0model}\\
 \label{eq:Fi}
 F_i &=& M^{eff} a^{eff} 
\eee

\noindent In this model, the inertial drag $F_i$ involves an effective mass of grains being set in motion and their typical acceleration, which are denoted by $M^{eff}$ and $a^{eff}$, respectively. 

To establish how these two parameters may be related to the plate size, plate depth and acceleration parameters, we consider the following elementary scenario involving the grain inertia and elasticity. Moving up the plate compresses a series of spring/mass elements. A spring element represents a grain to grain elastic contact with a stiffness $k = Ed$ and the mass element a grain mass $m$. Accordingly, accelerating the plate upward would generate an elastic wave propagating upward toward the free surface. Figure \ref{fig7} illustrates this process. Each grain/contact element acts as an harmonic oscillator which period is given by the collision time $t_c$. The acceleration wave thus propagates upward over a distance of one grain size $d$ at a speed scaling like $d/t_c$. The time $t_w$ for this elastic wave to reach the free surface is:

\be \label {eq:twave}
t_w = \frac{H}{d} t_c 
\ee

When the elastic wave reaches the surface, top grains move up freely realising the series of springs. Accordingly, the drag force should start relaxing then, at the latest. For relatively large acceleration times ($\tau>t_w$), the plate's acceleration is sustained at value close to $\frac{v_{\infty}}{\tau}$ throughout this process. As a result, all the grains above the plate are mobilised and contribute to the inertial resistance. We therefore express the corresponding effective mass and acceleration as:

\bee
a^{eff} &=&\frac{v_{\infty}}{\tau}\\
M^{eff} &=& \beta M_h
\eee

\noindent where $\beta$ is a dimensionless constant reflecting the extent of the zone of mobilised grain above the plate, which value is expected to be of the order of unity. 

For shorter acceleration times ($\tau<t_w$), the plate stops accelerating before the elastic wave reaches the free surface. As a result, not all the grains above the plate are mobilised before it stops accelerating. If the acceleration time $\tau$ of the plate becomes shorter than the collision time ($\tau<t_c$), even the first layer of grains would not have time to move before the plate stops accelerating. The fastest the first layer of grains can be mobilised and reached a velocity of $v_\infty$ is $t_c$. This defines an upper bound for the inertial drag, with an effective mass and acceleration given by:

\bee
a^{eff} &=& \frac{v_{\infty}}{t_c}\\
M^{eff} &=& M_h\frac{d}{H}
\eee

\noindent Accordingly, the inertial force in these two regimes can be expressed as $
F_i = \alpha v_{\infty}$ with:

\be 
\alpha = M_h 
  \begin{cases}
     \frac{\beta}{\tau} ,& \text{if } \tau \gg   t_w\\
     \frac{1}{ t_w},              & \text{if } \tau \ll   t_w
\end{cases}
\ee

\noindent We propose the following interpolation between these two regimes to obtain a continuous expression for the inertial drag force:

\bee
F_i &=& \alpha v_{\infty}  \label{eq:Fimodel}\\
\alpha &=& \frac{M_h}{t_w} \frac{1}{\frac{\tau}{\beta t_w}+1} \label{eq:alpha_predict}
\eee

\begin{figure*}
\centering
  \includegraphics[width=\textwidth]{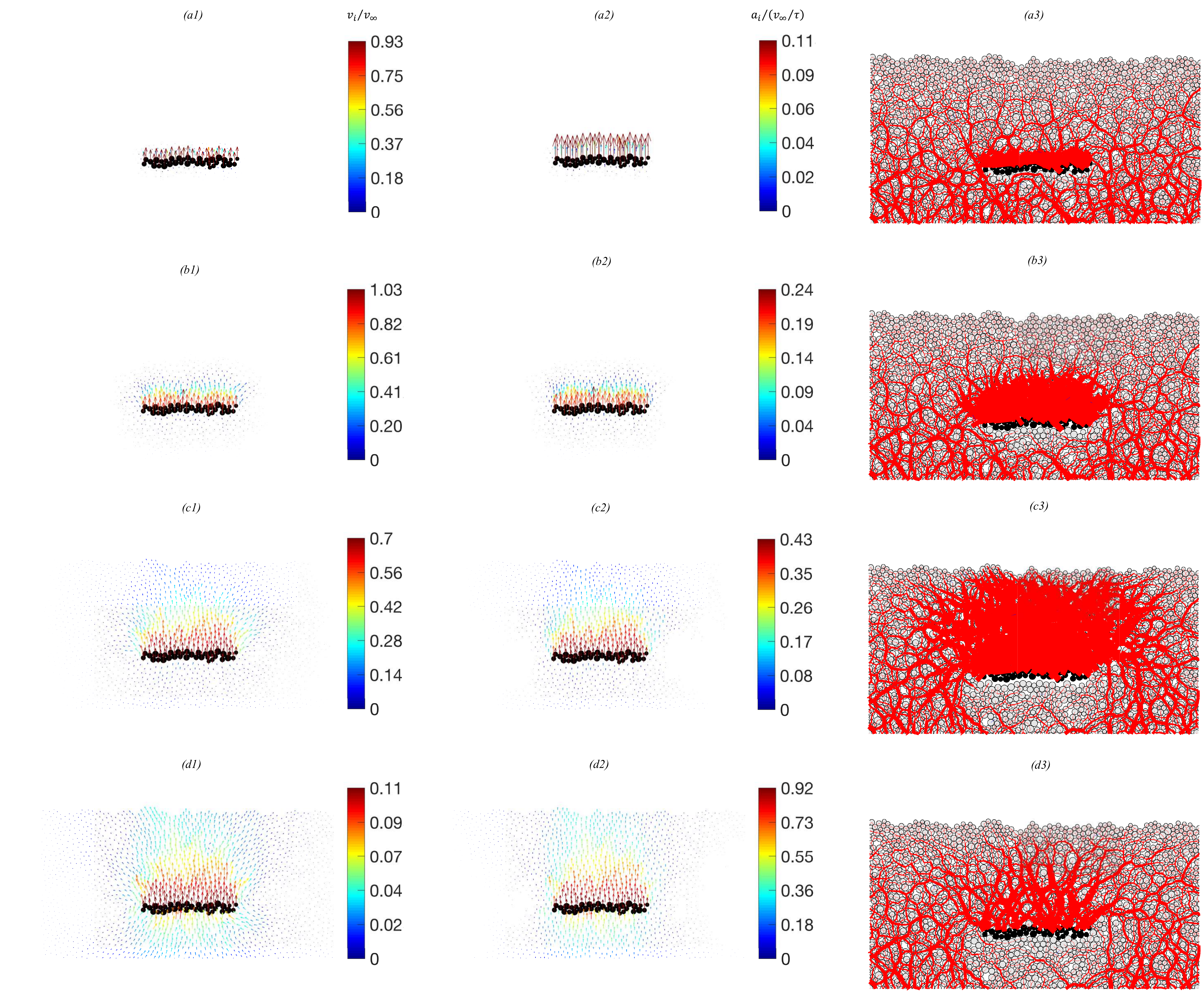}
\caption{Snapshots illustrating the grain mobilisation and contact compression when the maximum drag force is reached ($t=t_{peak}$). 
Rows (a-d) show uplift tests performed with $B=20d$, $H=B$ and $v_\infty = 3 \sqrt{gd}$ with differing acceleration times (top to bottom: $\tau/t_g = 10^{-3}$, $10^{-2}$, $10^{-1}$ and $2$).
 Grain vertical displacement (left column) and vertical acceleration (middle column) averaged from $t=0$ to $t_{peak}$. Normal contact force between grains at $t=t_{peak}$ (right column): red lines link pair of grains in contact, with a thickness proportional to the magnitude of the normal contact force, which is purely compressive as there is no inter-granular cohesion.
}
\label{Fig_8}      
\end{figure*}

 \begin{figure*}[!t]
\centering
  \includegraphics[width=0.80\textwidth]{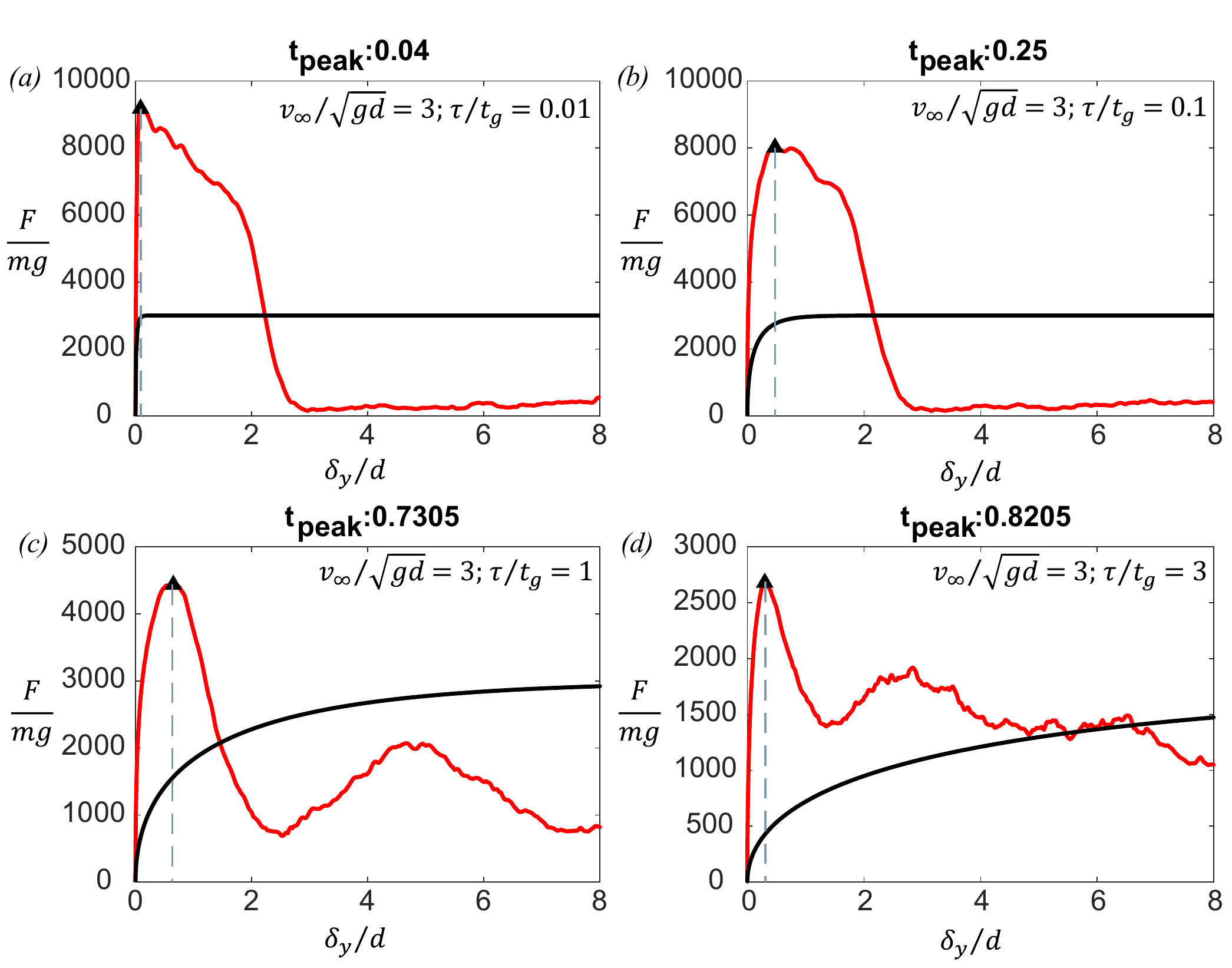}
\caption{Examples of drag force evolutions during uplift tests ($B = 30d$, $H/B = 1$, $v_\infty = 3\sqrt{gd}$) for different values of acceleration time $\tau$ showing different post-peak drag relaxation. Red lines denote drag forces and black lines represent the prescribed plate velocity $v(\delta_y(t))$, according to (\ref{eq:Vinfi}). Markers indicate the maximum drag force. Times $t_{peak}$ at which the maximum drag is reached are indicated in unit $t_g$.}
\label{Fig_9}      
\end{figure*}
 
\subsection{Assessing the elasto-inertial drag model}

The elasto-inertial drag model introduced in the previous section relies on a series of assumed physical processes, and leads to a prediction for the scaling of  the parameter $\alpha$ given by (\ref{eq:alpha_predict}). We use here the numerical results to assess the validity of these physical processes and scaling. 

\subsubsection{Scaling of $\alpha$}
Figure \ref{Fig_6}c compares the measured slopes $\alpha$ with the model prediction in Eq. (\ref{eq:alpha_predict}). When plotting the normalised slope $\alpha t_w/M_h$ as a function of the normalised acceleration time $\tau/t_w$, all numerical data obtained for different pate sizes $B$, different plate embedments $H$ and different grain stiffnesses $E$ collapse onto a single curve.
The prediction of the model  in equation (\ref{eq:alpha_predict}) quantitatively captures this curve in all regimes ($\tau<t_w$ and $\tau>t_w$) using a value $\beta=2$ as sole fitting parameter. This supports the validity of the final expression of the elasto-inertial drag.

\subsubsection{Partial/full mobilisation of grains above the plate}

Figure \ref{Fig_8} illustrates the contact forces and the grain displacements in the granular packing when the maximum drag force is reached. Grain displacements are analysed \textit{via} the average grain velocity and acceleration defined by:

\bee
 v_i &=& \frac{ x_i(t_{peak})-x_i(t=0)  }{t_{peak}} \\
 a_i &=& \frac{ x_i(t_{peak})-x_i(t=0)  }{t^2_{peak}/2}
\eee

\noindent where $x_i(t)$ is the position of a grain $i$ at time $t$, and $t_{peak}$ is the time at which the maximum drag force is reached. We opted to consider these time averaged values rather than the instantaneous velocities and accelerations because instantaneous values exhibit large fluctuations reflecting sudden and short lived grain rearrangements.

\begin{figure*}[ht]
\centering
\includegraphics[width=0.95\textwidth]{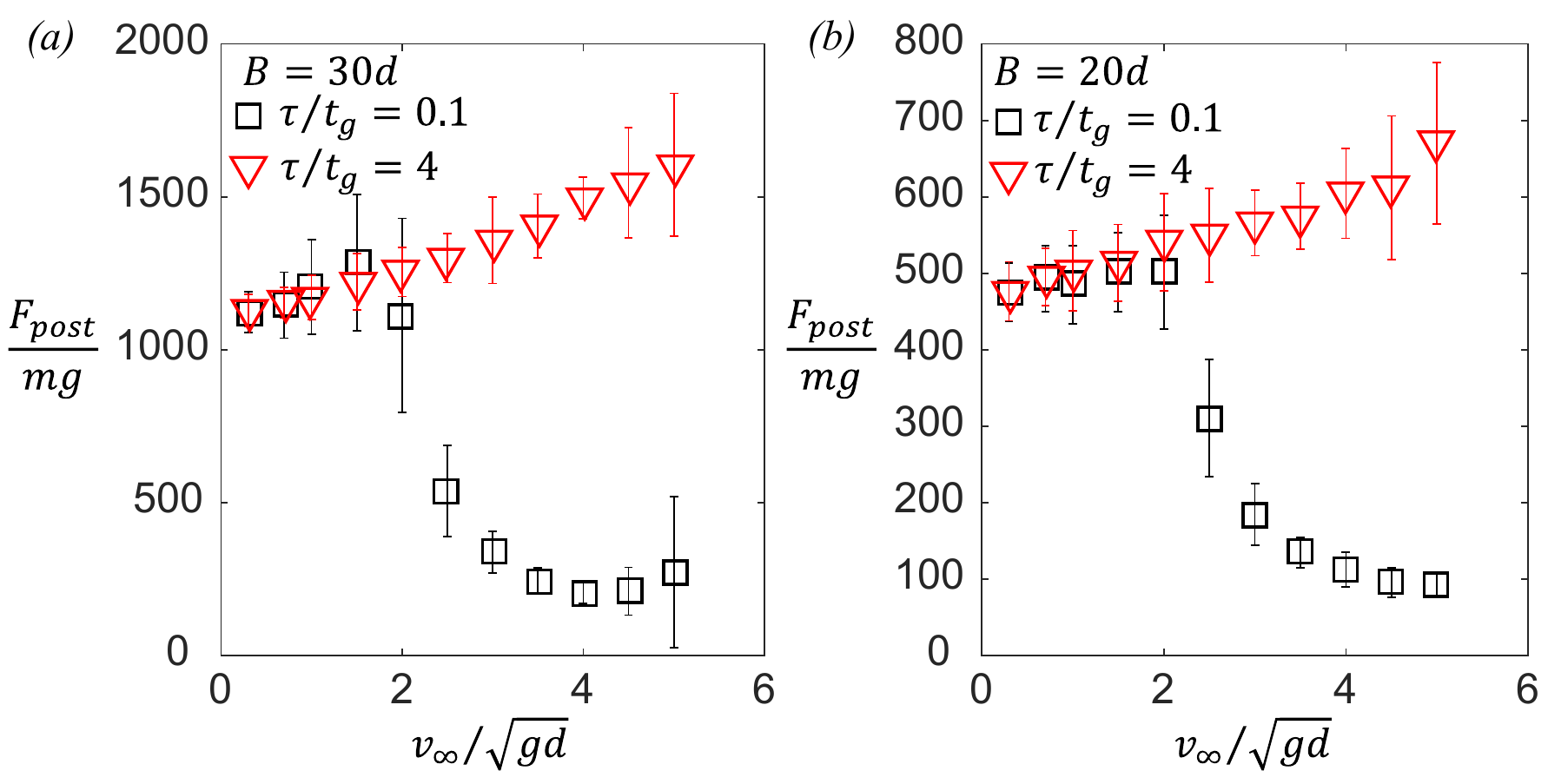}
\caption{Post-peak drag force $F_{post}$ (defined in Eq. \ref{eq:Fpost}) as a function of the plate ultimate velocity $v_\infty$. Tests shown here are performed with $H/B=1$, with different acceleration times and different plate sizes (see legends). Symbols and error bars show the average and standard deviation of $F_{post}$ obtained on a series of five tests with different realisations of the initial packing.}
\label{Fig_10}      
\end{figure*}

Comparative analysis of tests performed with different acceleration times qualitatively confirm the assumptions of the elasto-inertial model:
 
 \begin{itemize}
\item The first row on figure \ref{Fig_8} illustrates a test performed with a small acceleration time of $\tau = 10^{-3} t_g$. This is shorter than that the contact time $t_c$, which is $10^{-2} t_g$ in all tests presented on this figure. Most of the grains above the plate have not significantly moved when the maximum drag is reached, excepted for the first layer directly above the plate. Consistently, contact forces in these layers are highly compressed. The plate velocity has reached its final value $v_{\infty}$ and its averaged acceleration is lower than $v_\infty/\tau$, implying than the plate has finished accelerating before the maximum drag force is reached ($t_{peak}>\tau$). 
\item  At the other extreme, the last row on figure \ref{Fig_8} illustrates a test performed with a large acceleration time of $\tau = 2 t_g$ which is larger than the wave propagation time $t_w = 0.2 t_g$. All the grains located in the column above the plate, as well as some grains near this column, are mobilised when the maximum drag force is reached. The plate velocity is lower than $v_{\infty}$ and its acceleration is of the order of $v_\infty/\tau$, indicating that the plate is still accelerating when the maximum drag is reached ($t_{peak}<\tau$). The contact network exhibits some moderate compression from the plate to the free surface.
\item  The two central rows on figure \ref{Fig_8} show tests performed at intermediate values of $\tau$ larger than the collision time $t_c = 10^{-2}t_g$ but smaller than the wave propagation time $t_w=0.2 t_g$. They evidence that the elastic compression wave has not reached the free surface when the maximum drag is reached, and that only the first layers of grains that are the closest to the plate are mobilised, while the upper layers are not mobilised.
\end{itemize}

\begin{figure*}
\centering
  \includegraphics[width=\textwidth]{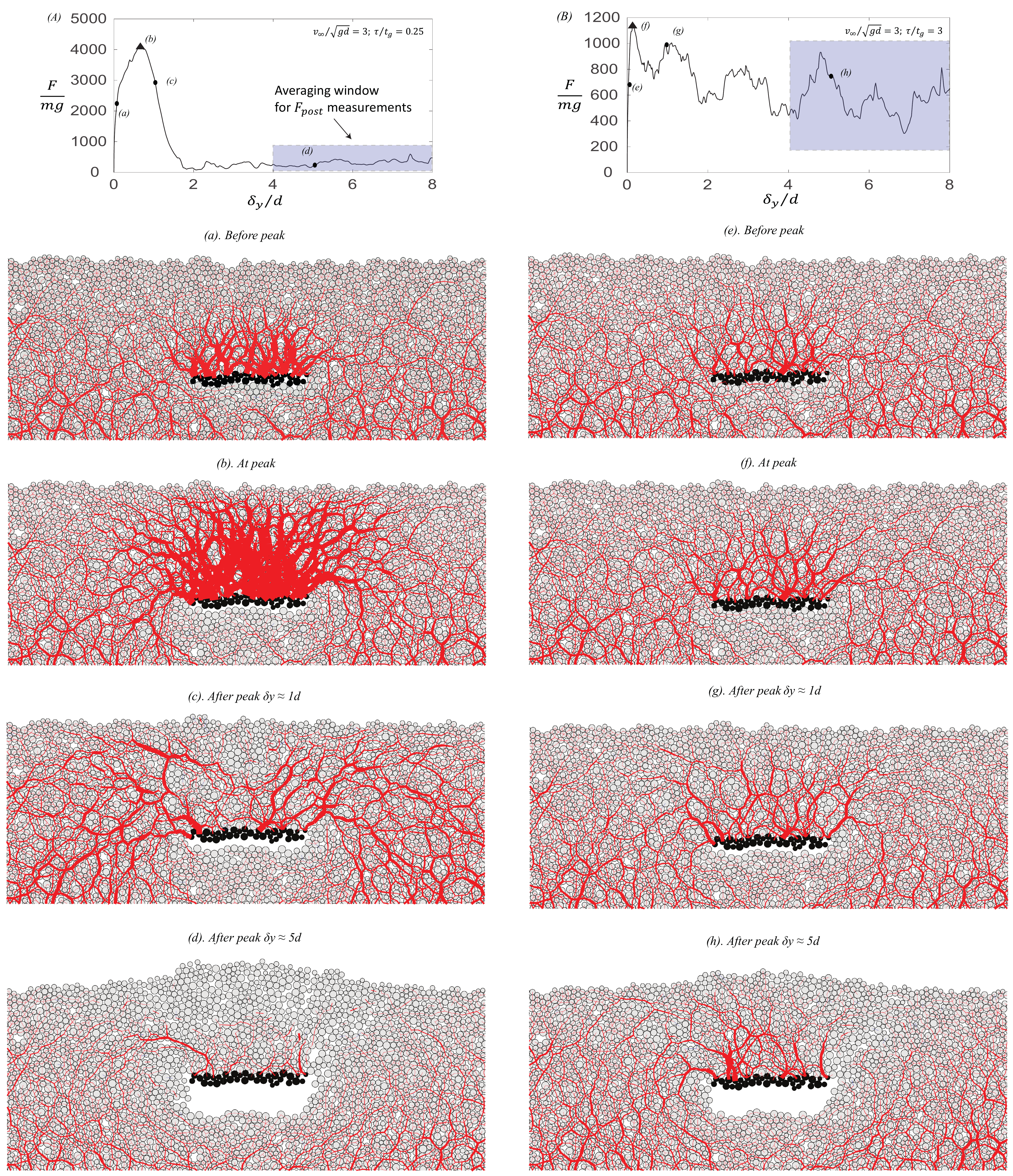}
\caption{Snapshot of contact network evolution during uplift. Right and left columns shows two tests performed with two different acceleration times $\tau = 0.25$ (left) and $\tau=3t_g$ (right); in both cases, $v_\infty = 3 \sqrt{gd}$, $H/d=20$, $H/B=1$. (A,B)
drag force evolution, indicating when the snapshots are taken. (a-h) corresponding force network: red lines denote contacts between the grains, with a width proportional to the normal contact force magnitude.}
\label{Fig_11}      
\end{figure*}

\section{Post peak drag relaxation}

The previous section have pointed out that grain elasticity and inertia influences the maximum drag force. This section focuses on the evolution of the drag force after the maximum drag is reached.

\subsection{Drag force evolution after peak}

Figure \ref{Fig_9} shows examples of drag force evolution during uplift for systems subjected to different acceleration times in the range $ 10^{-2} \leqslant \tau/t_g \leqslant 3$. For large acceleration times ($\tau>t_g$), the drag force gradually decays after the maximum is reached, with some fluctuations. In contrast, for small acceleration times  ($\tau<t_g$), the drag force sharply decays after the maximum is reached, to nearly zero. 

As a way to quantify this effect, we measured the value of the drag force after the peak. At large $\tau$, the post peak drag force fluctuates significantly with a period of about $d$ (Figure  \ref{Fig_9}b,c). We therefore consider the following average to compare tests under different conditions:

\be \label{eq:Fpost}
F_{post} = \frac{1}{4d}\int_{\delta_y=4d}^{8d} F(\delta_y) d \delta_y
\ee

\noindent which correspond to a small windows of displacement shortly after the maximum drag in all tests. 

Figure \ref{Fig_10} shows the values of the post peak drag force $F_{post}$ obtained for two plates, as a function of the ultimate velocity and acceleration time. 

For long acceleration times ($\tau=4t_g$),  post peak drag force linearly increases with the ultimate velocity $v_\infty$. This linear increase is similar to the maximum drag behaviour. This suggests that the plate is still accelerating after the maximum drag is reached, and that the post-peak drag is also enhanced by the inertia of the grains being accelerated in the packing.

For short acceleration times ($\tau=10^{-1}t_g$), post peak drag forces exhibit a similar linear increase with $v_\infty$ for $v_\infty \lesssim 2\sqrt{gd}$. At higher velocities, however, the post peak drag $F_{post}$ drops to a small value. This suggests that there is a mechanism that significantly weakens the granular packing, which only develops at high ultimate  velocities and short acceleration times.

Figure \ref{Fig_11} evidences this mechanism by showing the evolution of the contact network during two uplift tests performed with a high ultimate velocity ($v_\infty = 3 \sqrt{gd}$) and two different values of acceleration time ($\tau>t_g$ and $\tau<t_g$). At long acceleration time, the contact network above the plate is maintained before, when and after the maximum drag force is reached. In contrast, the test performed with a short acceleration time evidences a loss in contacts after the peak. The granular packing is then effectively fluidised and its resistance against the plate motion drops.

\subsection{Mechanisms of maximum drag force relaxation}

In the quasi-static regime ( $v_{\infty} \ll \sqrt{gd}$), drag force relaxation is driven by plastic deformation in the packing that contribute to relaxing some compressed contacts. These plastic deformations take the form of grain recirculation around the plate \cite{candelier2009creep,harich2011intruder,kolb2013rigid}. The criteria  $v_{\infty} \ll \sqrt{gd}$ can be interpreted as follows: grains can fall back under the plate by gravity quicker than the plate moves up. As a consequence, grain recirculation and its associated plastic deformations has enough time to continuously occur during the uplift.

Conversely, in the rate dependent regime ( $v_\infty \gg \sqrt{gd}$), grains do not have enough time to rearrange while the plate moves up. Figure \ref{Fig_11} evidences the formation of a gap under the plate as it moves up, and shows the upward deformation of the free surface resulting from the uplift of the packing above the plate. 

\begin{figure}
\centering
  \includegraphics[width=\columnwidth]{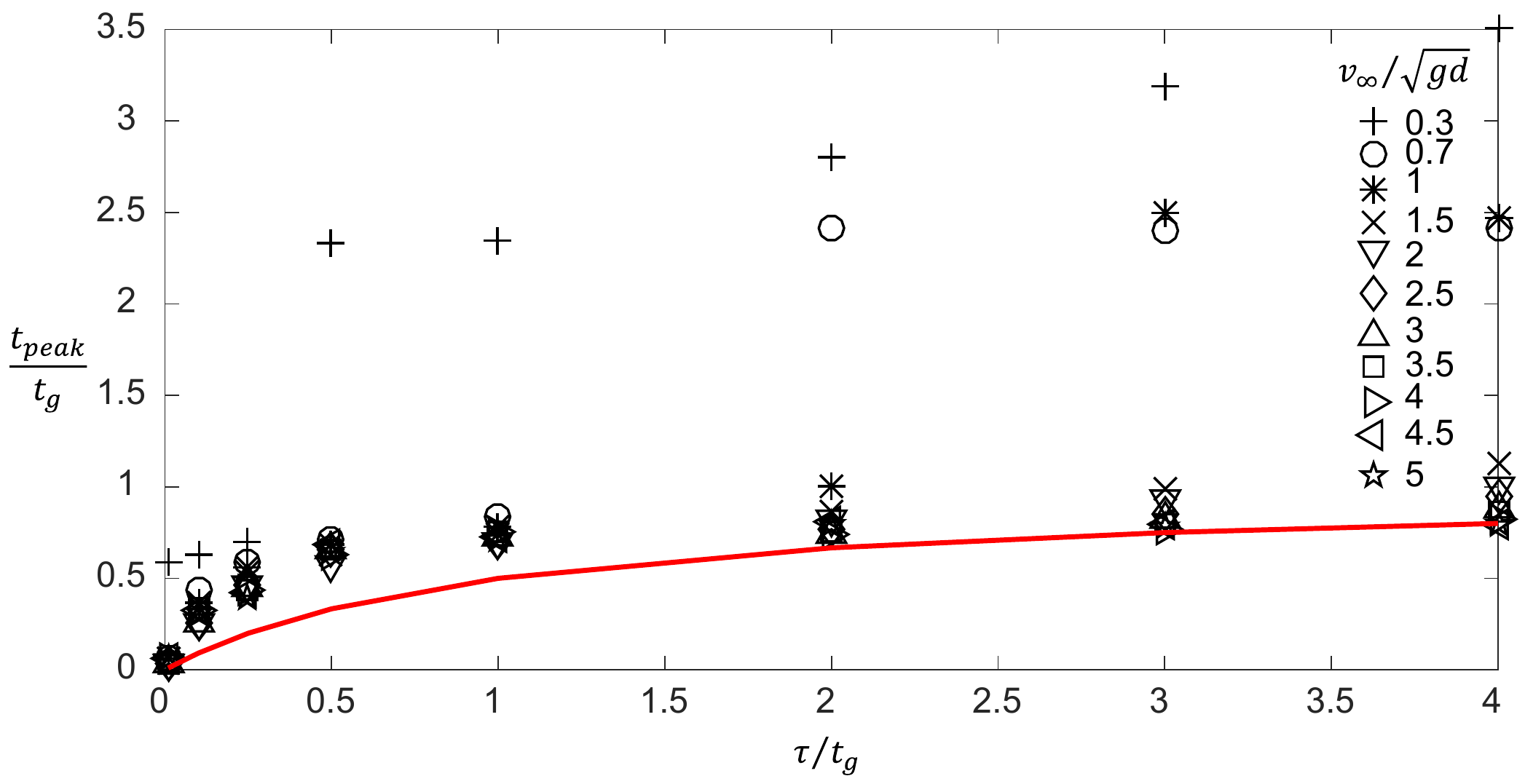}
\caption{Time $t_{peak}$ at which the drag force reaches its maximum and starts relaxing ($H/B=1$, $B/d=30$). Symbols represents test performed with different acceleration times and ultimate velocities (see legend). The red line shows the function $\frac{t_{peak}}{t_g} = \frac{\tau/t_g}{1+\tau/t_g}$ 
for visual reference, which behaves like $t_{peak} = \tau$ for ($\tau\ll t_g$) and like $t_{peak} = t_g$ for ($\tau \gg t_g$).
}
\label{Fig_12}      
\end{figure}

This suggests that the drag force starts to relax when grains can first rearrange by recirculating under the plate under the action of gravity. This mechanism implies that the maximum drag force is reached at $t_{peak}\approx t_g$. Figure \ref{Fig_12} shows that this is the case for long acceleration times ($\tau>t_g$), where $t_{peak}$ is larger than $\tau$. In contrast, at lower acceleration times ($\tau \ll t_g$), the maximum drag force relaxation corresponds to the end of the plate acceleration ($t_{peak} \approx \tau$), and is not controlled by grain recirculation around the plate.

\section{Conclusions}

This study points out that granular drag forces may be strongly affected by the acceleration of the moving object. 
We empirically evidenced this effect in a series of elementary uplift tests, and we rationalised it in terms of an elasto-inertial drag component resulting from the inertia of grains being mobilised in the packing.

The first finding is that the maximum drag force can always - at least for all the presented tests- be expressed in terms of a quasi-static component plus a dynamic component that is proportional to the final velocity of the object. This linear increase is expressed in (\ref{eq:F_v_emp}).  It defines the transition from a quasi-static to a rate-dependent drag regime occurring when the dynamic component becomes larger than the quasi-static component. We observed that this occurs when the ultimate plate velocity exceeds the grain free fall velocity scale $\sqrt{gd}$. Consequently, we propose that the quasi-static and rate-dependent regimes corresponds to whether or not grains have enough time to rearrange behind the moving plate to let it through the packing.  

The second finding is that the dynamic drag component results from an elasto-inertial process, by which grains in the packing are gradually being accelerated when the object is set into motion, with some delay. With the considered vertical uplift configuration, a full mobilisation is achieved when the plate acceleration is sustained long enough for the elasto-inertial compression wave it triggers to reach the surface.  For shorter acceleration times, we observed a partial mobilisation whereby only the layers the closest to the plate contribute to the inertial resistance. We introduced an inertial drag model based on this process that successfully captures the measured maximum drag forces. This model is expressed in Eqs. (\ref{eq:F0model}) and (\ref{eq:Fimodel}).

Finally, we observed that short-lived accelerations leads to an enhanced maximum drag force, but can lead to a  subsequent fluidisation of the packing. As a result, the drag force may drop to nearly zero after the maximum is reached.

The scope of this study is restricted to a particular mode of loading, the vertical uplift of a relatively shallow object. It is expected that similar inertial effect would arise with different mode of loading including vertical penetration, lateral ploughing, and motions a great depth \cite{albert1999slow,hilton2013drag,potiguar2013lift,guillard2013depth,seguin2019hysteresis}. At constant velocity, drag forces with these loadings are similar to that measured in vertical uplift; nonetheless, the zone of mobilised grains may be qualitatively different: it may not extend to the free surface and be localised around the object. How this would affect an elasto-inertial drag component remains to be measured and understood.

\bibliography{biblio}

\end{document}